\documentclass[pra,preprintnumbers,tightenlines,twocolumn,superscriptaddress]{revtex4-1}

\usepackage{amsmath}
\usepackage{amssymb}
\usepackage{graphicx}
\usepackage{psfrag}
\usepackage{color}
\usepackage[utf8]{inputenc}
\usepackage{ifthen}
\usepackage{listings}
\usepackage{placeins}
\usepackage{mathrsfs}
\usepackage{autobreak}
\allowdisplaybreaks

\usepackage{soul}

\usepackage{ulem}  
\graphicspath{{{./}}}

\begin{document}

\title{Photon-assisted Landau-Zener transitions in a periodically driven Rabi dimer coupled to a dissipative mode}

\author{Fulu Zheng}
\thanks{These two authors contributed equally}
\affiliation{Max-Planck-Institut für Physik komplexer Systeme, Nöthnitzer Strasse 38, D-01187 Dresden, Germany}

\author{Yuejun Shen}
\thanks{These two authors contributed equally}
\affiliation{Division of Materials Science, Nanyang Technological University, Singapore 639798, Singapore}

\author{Kewei Sun}
\affiliation{School of Science, Hangzhou Dianzi University, Hangzhou 310018, China}

\author{Yang Zhao}
\email{yzhao@ntu.edu.sg}
\affiliation{Division of Materials Science, Nanyang Technological University, Singapore 639798, Singapore}

\begin{abstract}
We investigate multiple photon-assisted Landau-Zener (LZ) transitions in a hybrid circuit quantum electrodynamics device in which each of two interacting transmission-line resonators is coupled to a qubit, and the qubits are driven by periodic driving fields and also coupled to a common phonon mode. The quantum state of the entire composite system is modeled using the multi-$\rm D_2$ {\it Ansatz} in combination with the time-dependent Dirac-Frenkel variational principle. Applying a sinusoidal driving field to one of the qubits, this device is an ideal platform to study the photon-assisted LZ transitions by comparing the dynamics of the two qubits. A series of interfering photon-assisted LZ transitions take place if the photon frequency is much smaller than the driving amplitude. Once the two energy scales are comparable, independent LZ transitions arise and a transition pathway is revealed using an energy diagram. It is found that both adiabatic and nonadiabatic transitions are involved in the dynamics. Used to model environmental effects on the LZ transitions, the common phonon mode coupled to the qubits allows for more available states to facilitate the LZ transitions. An analytical formula is obtained to estimate the short-time phonon population and produces results in reasonable agreement with numerical calculations. Equipped with the knowledge of the photon-assisted LZ transitions in the system, we can precisely manipulate the qubit state and successfully generate the qubit dynamics with a square-wave pattern by applying driving fields to both qubits, opening up new venues to manipulate the states of qubits and photons in quantum information devices and quantum computers.

\end{abstract}
\date{\today}

\maketitle

\section{introduction}

Nonadiabatic transitions represent one of the fundamental processes underlying photophysical and photochemical phenomena in molecules, molecular aggregates, and solids \cite{Crespo_Otero_CR_2018,Nakamura_book_2002}. Proposed by Landau, Zener, Stueckelberg, and Majorana independently in 1932, the Landau-Zener (LZ) model is the pioneering theoretical paradigm to investigate nonadiabatic transitions \cite{Landau_ZS_1932, Zener_PRSL_1932,Stueckelberg_1932, Majorana_1932}. The LZ model describes a driven two-level system with the energy spacing between two (diabatic) states being modulated as time evolves. As this energy separation changes sign, the diabatic states (i.e., eigenstates of the Hamiltonian in the absence of tunneling) undergo a {\it level crossing}, while the adiabatic states (i.e., eigenstates of the Hamiltonian in the presence of tunneling) encounter an {\it avoided crossing}. Characterizing both the nonadiabatic and the adiabatic transitions, the LZ model has been widely adopted to describe and understand fundamental physical problems in various fields, including atomic and molecular physics \cite{Niranjan_PRA_2020, Shevchenko_PR_2010, Salger_PRL_2007, Zhang_PRL_2018, Troiani_PRL_2017,Yu_PCCP_2014}, chemical physics \cite{Zhu_JCP_1997}, and quantum information science \cite{Matityahu_QI_2019,Petta_Science_2010, Quintana_PRL_2013}.

Thanks to their high flexibility and tunability, quantum electrodynamics (QED) devices are not only the  core units for quantum computing and quantum information \cite{Blais_PRA_2007, Billangeon_PRB_2015,Krantz_APR_2019, Wendin_RPP_2017, He_QIP_2020, Alqahtani_QIP_2020}, but also the ideal platforms for the realization of the LZ model \cite{Oliver_Science_2005, Wen_PRB_2020, Li_PSD_2020, Kervinen_PRL_2019}. For instance, Oliver {\it et al.} applied a periodic driving field to detune a superconducting flux qubit and the induced LZ transitions lead to multiphoton interference \cite{Oliver_Science_2005}. It is found that the strongly driven qubit has high stability and coherence, highlighting potential applications in quantum computing devices. It has also been proposed that LZ transitions can realize various functionalities in QED systems, such as quantum state preparation \cite{Saito_EPL_2006}, entanglement creation \cite{Wubs_PE_2007}, and even quantum bath calibration \cite{Wubs_PRL_2006}. In those proposals, a single driven qubit is coupled to a single or multiple oscillators. Environmental effects on the LZ transitions in these devices have also been systematically investigated by H\"anggi and coworkers \cite{Wubs_PRL_2006, Zueco_NJP_2008, Saito_PRB_2007}.

Beyond single qubit QED systems, hybrid circuit QED devices have also been proposed and fabricated recently to unveil the many-body quantum dynamics \cite{Koch_PRA_2010, Raftery_PRX_2014, Georgescu_RMP_2014}. For instance, Schmidt {\it et al.} proposed a hybrid QED device consisting of two coupled cavities with each containing a single qubit \cite{Schmidt_PRB_2010}. This device was designed to investigate fundamental light-matter interactions. A delocalization-localization transition of photons in this system is demonstrated theoretically by treating the device as a Jaynes-Cummings dimer (JCD) \cite{Schmidt_PRB_2010}. In 2014, they fabricated such a QED device, finding photon dynamics in the device to be consistent with their theoretical predictions \cite{Raftery_PRX_2014}. In order to better characterize the phenomena with ultrastrong qubit-photon coupling,
Hwang {\it et al.} extended the JCD model to a Rabi dimer model where the rotating-wave approximation is discarded. By calculating time-averaged photon imbalance under different qubit-photon coupling strengths and photon tunneling rates, they discovered three photonic phases \cite{Hwang_PRL_2016}. Recently the effects of a phonon environment and an external driving field on the dynamics of qubits and photons in the Rabi dimer have been elucidated in various parameter regimes using the time-dependent variational principle with the multiple Davydov D$_2$ {\it Ansatz} \cite{ZFL_AdP_2018, HZK_JCP_2019}. However, the LZ transitions and their susceptibility to environmental effects in this Rabi dimer system have not been well investigated, in contrast to systematically studied LZ transitions in single-qubit QED devices.

In this work, we apply a fully quantum mechanical method to study the LZ transitions in the Rabi dimer with the qubits driven by periodic driving fields and coupled to a common phonon mode. The quantum state of the hybrid system is described with the multiple Davydov D$_2$ {\it Ansatz} in the framework of the Dirac-Frenkel time-dependent variational principle. This method has been applied to investigate dissipative LZ transitions in a single qubit \cite{Huang_PRA_2018, Werther_JCP_2019}. By imposing a driving field to one of the two qubits, we observe a series of photon-assisted LZ transitions in the driven qubit. The transition pathway is revealed using the energy diagram of the coupled qubit and photon number states. Both adiabatic and nonadiabatic transitions are found to be involved in the dynamics. If the diagonal qubit-phonon coupling is turned on, more coupled states with small energy gaps are available to facilitate LZ transitions. In order to show the feasibility of quantum state engineering via LZ transitions, we generate qubit dynamics with a square-wave pattern by applying driving fields to both qubits.

The remainder of the paper is organized as follows. In Section \ref{methodology} we outline the theoretical framework of this study, including the system Hamiltonian, the multiple Davydov D$_2$ {\it Ansatz} and the time-dependent variational principle. The physical observables that we are interested in and the parameter configurations are also presented in this section. The main results and corresponding discussions are given in Section \ref{Numerical results and discussions}. Finally we conclude in Section \ref{conclusions}.

\section{methodology}
\label{methodology}
\subsection{System Hamiltonian}

\begin{figure}
  \centering
  \includegraphics[scale=0.4]{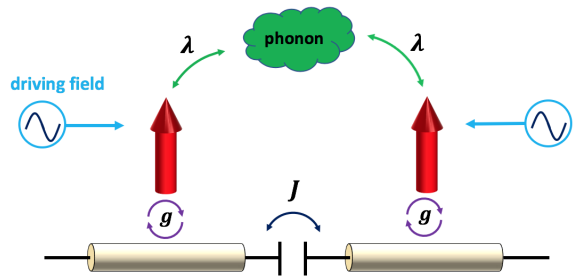}
  \caption{Schematic of the circuit QED device. Photons are able to hop between two transmission-line resonators with tunneling rate $J$. Left (right) qubit is off-diagonally coupled to left (right) resonator with coupling strength $g$. An external driving field may be applied to both qubits. The two qubits are coupled diagonally with a phonon mode with coupling strength $\lambda$. }
  \label{Fig1_model}
\end{figure}

We adopt a Rabi dimer to model the QED device consisting of two coupled transmission line resonators with each containing a qubit. As illustrated in Fig.~\ref{Fig1_model}, the qubits are driven by periodic driving fields and also coupled to a common phonon mode.
The Hamiltonian for the hybrid system is written as
\begin{equation}\label{eq:Htot}
 H = H_{\textrm{RD}} + H_\textrm{ph} + H_{\textrm{ph-q}},
\end{equation}
where $H_{\textrm{RD}}$, $H_\textrm{ph}$, $H_{\textrm{ph-q}}$ denote the Hamiltonians for the Rabi dimer, the phonon mode, and the qubit-phonon interaction, respectively.

In the Rabi dimer system ($ \hbar = 1 $), there are three terms in the Hamiltonian
\begin{equation}\label{eq:HRD}
   H_{\textrm{RD}}=H^{\textrm{Rabi}}_{\textrm{L}}+H^{\textrm{Rabi}}_{\textrm{R}}-J(a_{\textrm{L}}^{\dagger}a_{\textrm{R}}+a_{\textrm{R}}^{\dagger}a_{\textrm{L}}),
\end{equation}
and the photons can hop between the left (L) and right (R) resonators with a tunneling rate $J$. Within each resonator, the qubit-photon interaction is described with a Rabi model~\cite{Rabi_PR_1936,Rabi_PR_1937,Xie_JPA_2017,Braak_PRL_2011}
\begin{equation}\label{Hrabi}
  H_{i=\textrm{L}/\textrm{R}}^\textrm{Rabi} = \frac{F_{i}}{2}\cos(\Omega_{i}t+\Phi_{i}) \sigma_{z}^{i} + \omega_{i} a_{i}^{\dagger} a_{i} - g_{i} ( a_{i}^{\dagger} + a_{i} ) \sigma_{x}^{i}.\nonumber
\end{equation}
In contrast to the constant energy spacings of the two qubits in Ref.~\cite{ZFL_AdP_2018}, here we make them tunable by imposing a periodic harmonic driving field $F_{i}\cos(\Omega_{i}t+\Phi_{i})$ independently on each of the two qubits. In the $i$th ($i = \rm{L}, \rm{R}$) Rabi system, a photon mode with frequency $\omega_{i}$ is coupled to a qubit with an interaction strength $g_{i}$. $\sigma_{x}^{i}$ and $\sigma_{z}^{i}$ are the usual Pauli matrices, and $a_{i}$ ($a_{i}^{\dagger}$) is the annihilation (creation) operator of the $i$th photon mode.
We assume the two photon modes have identical frequencies, i.e., $\omega_{\textrm{L}}=\omega_{\textrm{R}}=\omega_{\textrm{r}}$, and are coupled to the corresponding qubit with the same coupling strength $g_{\textrm{L}}=g_{\textrm{R}}=g$.

The environmental effects on the Rabi dimer system are modeled by coupling a quantum harmonic oscillator
\begin{equation}\label{Hb}
  H_\textrm{ph}= \omega_{\textrm{ph}} b^{\dagger} b
\end{equation}
to the two qubits via the interaction Hamiltonian
\begin{equation}\label{Hbq}
  H_{\textrm{ph-q}}=\lambda (b^{\dagger}+b)(\sigma_{z}^{\textrm{L}}+\sigma_{z}^{\textrm{R}}),
\end{equation}
where $b$ ($b^{\dagger}$) is the annihilation (creation) operator of the single phonon mode with frequency $\omega_{\textrm{ph}}$, and $\lambda$ characterizes the qubit-phonon coupling strength.

If the energies corresponding to frequencies of the photon and phonon modes are higher than the thermal energy $k_BT$, the oscillators are thermally inactive, and thus the dynamics affected by the phonon is temperature independent in a wide temperature range \cite{Chi_Nature_2004,Wallraff_Nature_2004}. QED devices typically operate at extremely low temperatures \cite{Schmidt_AdP_2013}. Therefore, we set $T=0$ in the current study. It is straightforward to include the temperature effects in this methodology by applying Monte Carlo importance sampling \cite{WangLu_JCP_2017} and the thermo-field dynamics approach \cite{Chen_JCP_2017}. Especially, Werther {\it et al.} \cite{Werther_JCP_2019} have recently developed a generalized multi-D$_{2}$ {\it Ansatz} for displaced number states to investigate the LZ transition via number-state excitation. This method, based on the expansion of number states in terms of coherent states, can potentially capture the temperature effects in a numerically efficient manner. We believe that it can also be extended to our current model. Simulations at finite temperatures will be performed in future studies.

\subsection{The Multi-D$_2$ {\it Ansatz}}

The original Davydov {\it Ansätze} (solitons) were proposed in the 1970s to describe the solitary exciton in molecular systems \cite{Davydov_PSS_1973, Davydov_JTB_1973, Davydov_book_1985}. These simple {\it Ansätze} only provide approximate descriptions for a quantum system. For instance, a single Davydov D$_2$ trial state evaluated via the variational principle produces the same dynamics as that from the Ehrenfest approximation \cite{Huang_PCCP_2017}. Beyond the single versions of the Davydov {\it Ansatz}, Zhao and co-workers have proposed the multiple Davydov {\it Ansätze} \cite{Zhou_PRB_2014, Zhou_JCP_2015}, which are linear superpositions of the single Davydov trial states \cite{Zhao_JCP_2012,Zhao_JCP_1997}. These multiple Davydov {\it Ansätze} in principle can produce an exact solution to the Schrödinger equation in the limit of large multiplicities \cite{Zhou_PRB_2014, Zhou_JCP_2015, WangLu_JCP_2016, WangLu_JCP_2017, Werther_JCP_2019, Zhou_JPCA_2016}. The numerical accuracy and efficiency of the multiple Davydov {\it Ansätze} have been extensively verified in describing the states of various many-body systems \cite{Zhou_JCP_2015, WangLu_JCP_2016, WangLu_JCP_2017, Werther_JCP_2019, ZFL_AdP_2018, Huang_PRA_2018, Chen_CP_2018, HZK_JCP_2019}.

Although both the multiple Davydov D$_1$ (multi-D$_1$) and the multiple Davydov D$_2$ (multi-D$_2$) {\it Ansätze} are numerically exact with sufficiently large multiplicities, a suitable version for a specific problem has to be carefully chosen according to specific Hamiltonian constructs and parameter configurations. It has been found from extensive studies that the multi-D$_1$ {\it Ansatz} exhibits excellent performance for problems with diagonal system-bath coupling only, while the multi-D$_2$ {\it Ansatz} is suitable for tasks with off-diagonal system-bath coupling, despite that the multi-D$_2$ {\it Ansatz} has less variational parameters than the D$_1$ counterpart with a comparable multiplicity \cite{WangLu_JCP_2017, Werther_JCP_2019, Chen_CP_2018}.

The multi-${\rm D}_2$ {\it Ansatz} has been employed to study static properties and dynamic processes in a variety of toy models and realistic systems, demonstrating excellent numerical efficiency and accuracy in a broad parameter regime in the presence of both diagonal and off-diagonal system-bath couplings~\cite{Zhou_JCP_2015, Huang_PRA_2018, Zhou_JPCA_2016,Chen_JPCA_2017, Huang_PCCP_2017, Huang_JCPL_2017, Huang_AdP_2019}. Previous studies have also demonstrated the suitability and accuracy of the multi-D$_2$ {\it Ansatz} for similar driven QED systems \cite{HZK_JCP_2019}. In order to tackle both diagonal qubit-phonon and the off-diagonal qubit-photon interactions in the hybrid photon-qubit-phonon system in this work, we construct the following multi-${\rm D}_2$ {\it Ansatz} with multiplicity $M$ as a superposition of $M$ copies of the single Davydov D$_2$ {\it Ansatz}
\begin{eqnarray}\label{eq:MD2}
  |{\rm D}_{2}^{M}(t)\rangle &=& \sum_{n=1}^{M} \Big[ A_{n} (t)|\uparrow\uparrow\rangle + B_{n} (t) |\uparrow\downarrow\rangle + C_{n} (t) |\downarrow\uparrow\rangle \nonumber\\
  &&~~~~+ D_{n} (t) |\downarrow\downarrow\rangle \Big] \bigotimes |\mu_{n}\rangle_{\textrm{L}}|\nu_{n}\rangle_{\textrm{R}} |\eta_{n}\rangle_{\textrm{ph}}
\end{eqnarray}
to describe the quantum state of the composite photon-qubit-phonon system. Here,
$|\uparrow \downarrow \rangle=| \uparrow \rangle_{\textrm{L}} \otimes | \downarrow \rangle_{\textrm{R}}$ represents a state of the two qubits with $| \uparrow \rangle$ $(| \downarrow \rangle)$ indicating the up (down) state of a qubit. $|\mu_{n}\rangle_{\textrm{L}}$ and $|\nu_{n}\rangle_{\textrm{R}}$ are coherent states of the photon modes in the two resonators
\begin{eqnarray}
  |\mu_{n}\rangle_{\textrm{L}} & = & \exp\left[\mu_{n} (t) a_{\textrm{L}}^{\dagger}-\mu_{n}^{\ast} (t) a_{\textrm{L}}\right]|0\rangle_{\textrm{L}},\\
  |\nu_{n}\rangle_{\textrm{R}} & = & \exp\left[\nu_{n} (t) a_{\textrm{R}}^{\dagger}-\nu_{n}^{\ast} (t) a_{\textrm{R}}\right]|0\rangle_{\textrm{R}},
\end{eqnarray}
where $|0\rangle_{\textrm{L}(\textrm{R})}$ is the vacuum state of the photon mode in the left (right) resonator. The state of the phonon mode is also represented using a coherent state
\begin{equation}\label{eta}
  |\eta_{n}\rangle_{\textrm{ph}} = \exp \left[ \eta_{n} (t)  b^{\dagger}-\eta_{n}^{\ast} (t) b \right] |0\rangle_{\textrm{ph}}
\end{equation}
with a phonon vacuum state $|0\rangle_{\textrm{ph}}$. 
The state of the system Eq.~(\ref{eq:MD2}) is constructed once the time-dependent variational parameters, i.e., $A_{n}(t)$, $B_{n}(t)$, $C_{n}(t)$, $D_{n}(t)$, $\mu_{n}(t)$, $\nu_{n}(t)$, and $\eta_{n}(t)$, are determined via the variational principle as demonstrated below.
$A_{n}$ is the probability amplitude in the state $|\uparrow\uparrow\rangle|\mu_{n}\rangle_{\textrm{L}}|\nu_{n}\rangle_{\textrm{R}}|\eta_{n}\rangle_{\textrm{ph}}$, $\mu_{n}$ ($\nu_{n}$) is the displacement of the left (right) photon mode, and $\eta_{n}$ is the displacement of the phonon mode.

\subsection{The time-dependent variational principle}

The evolution of the variational parameters is governed by the equations of motion of the parameters, which are derived using the Dirac-Frenkel time-dependent variational principle
\begin{equation}\label{DiracFrenkel}
  \frac{d}{dt} \bigg( \frac{\partial \mathscr{L}}{ \partial \dot{\alpha}^{*}_{n}} \bigg) - \frac{\partial \mathscr{L}}{ \partial \alpha^{*}_{n}} =0.
\end{equation}
Here $\alpha_{n}$ are the variational parameters, i.e., $A_{n}(t)$, $B_{n}(t)$, $C_{n}(t)$, $D_{n}(t)$, $\mu_{n}(t)$, $\nu_{n}(t)$, and $\eta_{n}(t)$, and $\mathscr{L}$ is the Lagrangian given by
\begin{equation}\label{Lagrangian}
  \mathscr{L} = \frac{i}{2} \langle {\rm D}_{2}^{M}(t) | \frac{\overrightarrow{\partial}}{\partial t} - \frac{\overleftarrow{\partial}}{\partial t} | {\rm D}_{2}^{M}(t) \rangle - \langle {\rm D}_{2}^{M}(t) | H | {\rm D}_{2}^{M}(t) \rangle.
\end{equation}
The evolution equations for all variational parameters are given in Appendix~\ref{Equations of Motion}.

\subsection{Observables}

Employing the Dirac-Frenkel time-dependent variational principle with the multi-D$_2$ {\it Ansatz}, we investigate the dynamical processes in the hybrid system by monitoring the real-time dynamics of photon and phonon numbers, the qubit state, and the LZ transition probability. The time evolution of photon numbers in two resonators is given by
\begin{eqnarray}
  N_{\textrm{L}}(t) & = & \langle{\rm D}_{2}^{M}(t)| a_{\textrm{L}}^{\dagger} a_{\textrm{L}} |{\rm D}_{2}^{M}(t) \rangle \nonumber\\
  & = & \sum_{l,n}^{M} \Big[ A_{l}^{\ast}(t) A_{n}(t)  + B_{l}^{\ast}(t) B_{n}(t)  + C_{l}^{\ast}(t) C_{n}(t) \nonumber\\
  &&~~~~~~~ + D_{l}^{\ast}(t) D_{n}(t) \Big] \mu_{l}^{\ast}(t) \mu_{n}(t)  S_{ln}(t) , \\
  N_{\textrm{R}}(t) & = & \langle{\rm D}_{2}^{M}(t)|a_{\textrm{R}}^{\dagger}a_{\textrm{R}}|{\rm D}_{2}^{M}(t) \rangle \nonumber \\
  & = & \sum_{l,n}^{M} \Big[ A_{l}^{\ast}(t) A_{n}(t)  + B_{l}^{\ast}(t) B_{n}(t)  + C_{l}^{\ast}(t) C_{n}(t) \nonumber\\
  &&~~~~~~~ + D_{l}^{\ast}(t) D_{n}(t) \Big] \nu_{l}^{\ast}(t) \nu_{n}(t)  S_{ln}(t) ,
\end{eqnarray}
with $S_{ln} (t)$ being the Debye-Waller factor
\begin{eqnarray}
  S_{ln} & = &  \exp\left[\mu_{l}^{\ast}(t) \mu_{n}(t)-\frac{1}{2} |\mu_{l}(t)|^2-\frac{1}{2}|\mu_{n}(t)|^2\right] \cdot \nonumber\\
  && \exp\left[ \nu_{l}^{\ast}(t) \nu_{n}(t) - \frac{1}{2} |\nu_{l}(t)|^2 - \frac{1}{2} |\nu_{n}(t)|^2 \right] \cdot \nonumber\\
  &&\exp \left[   \eta_{l}^{\ast}(t) \eta_{n}(t) - \frac{1}{2} |\eta_{l}(t)|^2 - \frac{1}{2} |\eta_{n}(t)|^2  \right].
\end{eqnarray}
The time evolution of the total photon number is ${N}(t)=N_{\textrm{L}}(t)+N_{\textrm{R}}(t)$. In order to characterize photon localization and delocalization, we also analyze the photon imbalance $Z(t)=N_{\textrm{L}}(t)-N_{\textrm{R}}(t)$.

In addition to the photon numbers in two resonators, we also monitor the population dynamics on the number states of the photon modes. By projecting the total wave packet on photon number states $|n_{\rm L} \rangle$ and $|n_{\rm R} \rangle$, the photon populations on these states are expressed as
\begin{eqnarray}
  &&|\langle n_{\rm L} |{\rm D}_{2}^{M}(t)\rangle |^2 \nonumber\\
  &=& \sum_{l,n}^{M} \exp\left[-\frac{|\mu_{l}(t)|^2 |\mu_{n}(t)|^2}{2}\right] \times \frac{\left[\mu_{l}^{\ast}(t)\mu_{n}(t)\right]^{n_L}}{n_L!} \nonumber \\     
  &&\Big[ A_{l}^{\ast}(t) A_{n}(t)  + B_{l}^{\ast}(t) B_{n}(t)  + C_{l}^{\ast}(t) C_{n}(t) + D_{l}^{\ast}(t) D_{n}(t) \Big] \nonumber\\
 &&\times \exp\left[\nu_{l}^{\ast}(t) \nu_{n}(t)-\frac{1}{2} |\nu_{l}(t)|^2-\frac{1}{2}|\nu_{n}(t)|^2\right] \nonumber\\
  &&\times \exp \left[   \eta_{l}^{\ast}(t) \eta_{n}(t) - \frac{1}{2} |\eta_{l}(t)|^2 - \frac{1}{2} |\eta_{n}(t)|^2  \right]
\end{eqnarray}

\begin{eqnarray}
  &&|\langle n_{\rm R} |{\rm D}_{2}^{M}(t)\rangle |^2 \nonumber\\
  &=& \sum_{l,n}^{M} \exp\left[-\frac{|\nu_{l}(t)|^2 |\nu_{n}(t)|^2}{2}\right] \times \frac{\left[\nu_{l}^{\ast}(t)\nu_{n}(t)\right]^{n_R}}{n_R!} \nonumber \\     
  &&\Big[ A_{l}^{\ast}(t) A_{n}(t)  + B_{l}^{\ast}(t) B_{n}(t)  + C_{l}^{\ast}(t) C_{n}(t) + D_{l}^{\ast}(t) D_{n}(t) \Big] \nonumber \\
  &&\times \exp\left[\mu_{l}^{\ast}(t) \mu_{n}(t)-\frac{1}{2} |\mu_{l}(t)|^2-\frac{1}{2}|\mu_{n}(t)|^2\right] \nonumber\\
  &&\times \exp \left[   \eta_{l}^{\ast}(t) \eta_{n}(t) - \frac{1}{2} |\eta_{l}(t)|^2 - \frac{1}{2} |\eta_{n}(t)|^2  \right]
\end{eqnarray}

Along with the photon dynamics, the time evolution of the qubit states is recorded in the simulations by measuring the qubit polarization at every time point via
\begin{eqnarray}
  \langle\sigma_{z}^{\textrm{L}}(t)\rangle & = & \langle{\rm D}_{2}^{M}(t)|\sigma_{z}^{\textrm{L}}|{\rm D}_{2}^{M}(t) \rangle \nonumber \\
  & = & \sum_{l,n}^{M} \Big[ A_{l}^{\ast}(t)A_{n}(t) + B_{l}^{\ast}(t)B_{n}(t) \nonumber\\
  &&- C_{l}^{\ast}(t)C_{n}(t) - D_{l}^{\ast}(t)D_{n}(t) \Big] S_{ln}(t),\\
  \langle\sigma_{z}^{\textrm{R}}(t)\rangle & = & \langle{\rm D}_{2}^{M}(t)|\sigma_{z}^{\textrm{R}}|{\rm D}_{2}^{M}(t) \rangle \nonumber \\
  & = & \sum_{l,n}^{M} \Big[ A_{l}^{\ast}(t)A_{n}(t) - B_{l}^{\ast}(t)B_{n}(t) \nonumber\\
  &&+ C_{l}^{\ast}(t)C_{n}(t) - D_{l}^{\ast}(t)D_{n}(t) \Big] S_{ln}(t).
\end{eqnarray}

Assuming that both qubits are initiated from their down states, we also measure the LZ transition probability of the qubits flipping to their up states via
\begin{eqnarray}
  P_{\rm LZ}^{\textrm{L}}(t) & = &{\Big| \langle\uparrow_{\rm L} |{\rm D}_{2}^{M}(t)\rangle \Big|}^2 \nonumber \\
  & = & \sum_{l,n}^{M} \Big[ A_{l}^{\ast}(t)A_{n}(t) + B_{l}^{\ast}(t)B_{n}(t) \Big] S_{ln}(t),  \nonumber \\ 
  P_{\rm LZ}^{\textrm{R}}(t) & = &{\Big| \langle\uparrow_{\rm R} |{\rm D}_{2}^{M}(t)\rangle \Big|}^2 \nonumber \\
  & = & \sum_{l,n}^{M} \Big[ A_{l}^{\ast}(t)A_{n}(t) + C_{l}^{\ast}(t)C_{n}(t) \Big] S_{ln}(t).  \nonumber \\
\end{eqnarray}

As demonstrated above, our fully quantum mechanical method based on the wave function of the whole system can explicitly produce the population dynamics of the phonon mode. Analogous to the photon population, the phonon population is calculated as
\begin{eqnarray}
  N_{\textrm{ph}}(t) &=& \langle{\rm D}_{2}^{M}(t)| b^{\dagger} b|{\rm D}_{2}^{M}(t) \rangle \nonumber \\
  &=& \sum_{l,n}^{M} \Big[ A_{l}^{\ast}(t) A_{n}(t) + B_{l}^{\ast}(t) B_{n}(t) + C_{l}^{\ast}(t) C_{n}(t) \nonumber\\
  &&~~~~~~~~+ D_{l}^{\ast}(t) D_{n}(t) \Big] \eta_{l}^{\ast}(t) \eta_{n}(t) S_{ln}(t).
\end{eqnarray}

\subsection{Parameter configurations and initial conditions}

In a bare resonant Rabi dimer given by Hamiltonian (\ref{eq:HRD}), where each of the two coupled photon modes interacts with a qubit individually and the photon frequency is the same as the qubit bias, the dynamics of the photons and the qubits are governed by the competitive effects of the qubit-photon coupling $g$ and the inter-resonator photon hopping rate $J$ \cite{Hwang_PRL_2016, ZFL_AdP_2018}. Hwang {\it et al.} constructed a phase diagram of the time-averaged photon imbalance $Z(t)$ with respect to $g$ and $J$ for this system, and discovered localized and delocalized photonic phases \cite{Hwang_PRL_2016}. It is known that the photon-assisted LZ transition probability is dependent on the average photon number in the resonator~\cite{Sun_PRA_2012} and a fast change of the average photon number during the transition time may lead to a large deviation from the standard LZ transition phenomena. To better understand the LZ transition in the Rabi dimer with periodic driving, we adopt a small photon tunneling rate $J=0.01~\omega_0$ and a relatively strong qubit-photon coupling strength $g=0.3~\omega_0$, leading to a photon localization in the bare resonant Rabi dimer system \cite{Hwang_PRL_2016, ZFL_AdP_2018}. Here $\omega_0$ is the unit frequency. In addition to realizing LZ transitions, the periodic driving fields applied to the qubits in the Rabi dimer can manipulate the states of the photons and the qubits. The state manipulation is demonstrated via an example with a parameter configuration of $J=0.075~\omega_0$ and $g=0.3~\omega_0$, which is located in the delocalized photonic phase in the bare resonant Rabi dimer.

The driving amplitude is set to be large to prevent the qubit flipping. Therefore, the off-diagonal qubit-photon coupling is not able to trap the photons in the resonators. As a result, in our work, the photon is always delocalized. Also, the driving frequency is chosen to be small to allow a slow sweep which facilitates adabatics transitions.

We prepare the dimer by pumping $N(0)=20$ photons into the left resonator and keeping the right resonator in a photon vacuum with $\mu_1(t=0)=\sqrt{20}$ and $\mu_{n\neq1}(t=0)=\nu_n(t=0)=0$. The phonon mode is initially in its vacuum state with $\eta_{n}(t=0)=0$. Both the left and right qubits are initially prepared in their down states with $A_n(t=0)=B_n(t=0)=C_n(t=0)=D_{n\neq1}(t=0)=0$ and $D_1(t=0)=1$. The convergence of our method has been proved in our previous work \cite{ZFL_AdP_2018}. Balancing between accuracy and computational cost, we choose multiplicity $M=6$.

\section{Results and discussion}
\label{Numerical results and discussions}

In order to investigate the LZ transitions in the system, we focus more on the total photon number and total energy change due to the following reasons. There are two ways to affect the photon number in the left and right resonators. Firstly, photons can hop between the two resonators. In our case, this is the major way of controlling the photon number. In an extreme case where $g=0$, i.e., without the qubit-photon coupling, the average photon number in the two resonators will follow a sinusoidal curve with a Josephson oscillation period of $T_J=2\pi/2J$ while the total photon number is conserved. Secondly, qubits may create or annihilate photons while at the same time flipping their states under the external field. This is where photon assisted LZ transitions take place. The total photon number is not conserved. Therefore, to understand the photon number change through the second channel, more attention will be paid to the total photon number change. The total energy is independent of the photon tunneling since the total photon number is conserved in the absence of qubit-photon coupling. However, it depends on the qubit-photon coupling (the second channel) as the energy may evolve according to the energy diagram.

We show our results in the following order. Firstly, we discuss the case of a strong driving field on the left qubit only and the low photon frequency in the absence of the phonon mode. Secondly, we switch to a case similar to the above case but with high frequency photons, which enables us to explicitly analyze the underlying photon-assisted LZ transitions. Third, we couple a dissipative phonon mode to the qubits and compare it with the previous results. At last, we manage to manipulate the qubit state and the photon dynamics in the absence of the phonon mode by tuning the external driving fields on two qubits.

\subsection{Only the left qubit is driven}
In this subsection, we discuss LZ transitions and photon dynamics in the Rabi dimer when only the left qubit is driven by a periodic field. As shown in Fig.~\ref{Fig2_large_driving_low_freq} (a), the driving fields are configured with following parameters $F_{\rm L}=20~\omega_0$, $\Omega_{\rm L}=0.05~\omega_0$,  $\Phi_{\rm L}=0$, $F_{\rm R}=0$, $\Omega_{\rm R}=0$, and $\Phi_{\rm R}=0$. We consider three cases in this part. In the first two cases, the qubits are coupled to photons only ($\lambda = 0$), either low frequency ($\omega_{\textrm{r}}=\omega_0$) or high frequency ($\omega_{\textrm{r}}=10~\omega_0$) photons. In the third case, the qubits are coupled a phonon mode ($\lambda = 0.2, 0.4$) as well as high frequency ($\omega_{\textrm{r}}=10~\omega_0$) photons.

\subsubsection{Qubits coupled to low frequency photons}

\begin{figure}
  \centering
  \includegraphics[scale=0.5]{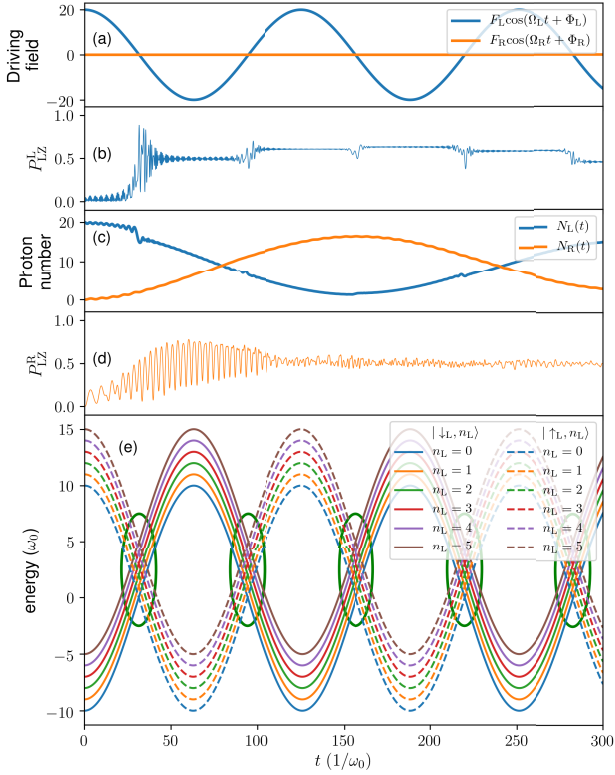}
  \caption{Dynamics of the Rabi dimer with the qubits coupled to photons with $\omega_{\textrm{r}}=\omega_0$, and driven by fields (a) parameterized with $F_{\rm L}=20~\omega_0$, $\Omega_{\rm L}=0.05~\omega_0$, $\Phi_{\rm L}=0$, $F_{\rm R}=0$, $\Omega_{\rm R}=0$, $\Phi_{\rm R}=0$. The LZ transition probabilities of the left and right qubits are illustrated in (b) and (d), respectively. Dynamics of the photon numbers in two resonators is shown in (c). Panel (e) presents the energy diagram of the coupled states of the left qubit and the number states of the photons in the left resonator. As the diagram is infinitely large, here we only show the coupled states of the left qubit and the photon number states with $n_{\rm L} = 0,1,...,5$. Since the coupling strength $g>0.1~\omega_{\textrm{r}}$, it is in the ultra-strong coupling regime. Therefore, high order LZ transitions can take place within the green circles.}
  \label{Fig2_large_driving_low_freq}
\end{figure}

Driven by the field shown as the blue curve in Fig.~\ref{Fig2_large_driving_low_freq} (a) and coupled to a photon mode with frequency $\omega_{\textrm{r}}=\omega_0$, the left qubit undergoes a series of LZ transitions as shown in Fig.~\ref{Fig2_large_driving_low_freq} (b) when the driving field passes through zero and an avoid crossing occurs. In contrast, the undriven right qubit just flips frequently due to the off-diagonal qubit-photon coupling as depicted in Fig.~\ref{Fig2_large_driving_low_freq} (d). Comparing the photon number dynamics in the left resonator shown in Fig.~\ref{Fig2_large_driving_low_freq} (c) and the LZ transition probability of the left qubit, we can clearly see that the LZ transitions take place in company of sudden changes on photon numbers, indicating they are indeed photon-assisted transitions.

In order to understand the LZ transitions of the left qubit, we calculate the time evolution of the discrete energies of the hybrid states $|\downarrow_{\rm L}(\uparrow_{\rm L}),n_{\rm L} \rangle$, where $|\downarrow_{\rm L} (\uparrow_{\rm L}) \rangle$ is the left qubit state and $| n_{\rm L} \rangle$ is the number state of the left photon mode. The energy diagram is illustrated in Fig.~\ref{Fig2_large_driving_low_freq} (e). For simplicity, we only show 6 up states and 6 down states of the left qubit, i.e., $|\downarrow_{\rm L} (\uparrow_{\rm L}),n_{\rm L}\rangle, n_{\rm L} = 0, 1, 2, 3, 4, 5$. The energy gap between two neighboring states with the same qubit state is always equal to $\omega_r$. Note that here we use the discrete photon number state instead of the continuous coherent state in the energy diagram. Even though the photons in resonators are prepared in coherent states, they can always be decomposed into infinite series of number states. To be specific, by decomposing the coherent states in Eq.~(\ref{eq:MD2}), the system state without phonon can always be written as
\begin{eqnarray} \nonumber
|\widetilde{\rm D}_{2}^{M}(t)\rangle &=& \sum_{n=1}^{M} \exp\Big(-\frac{|\mu_{n} |^{2}+|\nu_{n}|^{2}} {2}\Big)\Big[ A_{n} (t)|\uparrow\uparrow\rangle \\
&&~~~~ + B_{n} (t) |\uparrow\downarrow\rangle + C_{n} (t) |\downarrow\uparrow\rangle \nonumber + D_{n} (t) |\downarrow\downarrow\rangle \Big] \\
&&~~~~\bigotimes \sum_{i=0}^{\infty} \frac{\mu_{n}^{i}}{\sqrt {i!}}|i \rangle_{\textrm{L}} \sum_{j=0}^{\infty } \frac{{\nu_{n}^{j}}} {\sqrt {j!}}|j\rangle_{\textrm{R}}.
\end{eqnarray}

The evolution of every number state follows the energy diagram. Therefore, the photon coherent state evolves in the same manner since it is a superposition of the number states. When $\omega_{\textrm{r}}\ll F_{\rm L}$, there are a large number of energy cross points in a short time interval when the up-state and down-state clusters sweep across. Moreover, $g>0.1~\omega_{\textrm{r}}$ implies that it is in the ultra-strong-coupling regime. Therefore, multiple high order LZ transitions may take place in the green circles in Fig.~\ref{Fig2_large_driving_low_freq} (e). To compare with the linear LZ transition problem, we estimate the sweep rate $v_L$ using the time derivative of the driving field when the state energies cross, and get $v_{\rm L} = F_{\rm L} \Omega_{\rm L} = {\omega_0}^2$.

Now we are in a good position to analyze the LZ transitions in the left Rabi monomer. As we can see from Figs.~\ref{Fig2_large_driving_low_freq} (b) and \ref{Fig2_large_driving_low_freq} (e), there are stepwise changes of $P_{\rm LZ}^{\rm L}$ at the time when a large number of energy crossings take place. We examine the first LZ transition for $30/\omega_0<t<40/\omega_0$ as an example. The average photon number in the left resonator during transition is $N_{\rm L}(t)\approx 17$. As a result, the relaxation time of the LZ transition $\tau_{\rm LZ} \sim 2g\sqrt{N_{\rm L}(t)+1}/v_{\rm L}$ is longer than the time gap between the two LZ transitions $t_{c} \sim \omega_{\textrm{r}}/v_{\rm L}$. Therefore, the series of LZ transitions are not independent and hence a beating pattern instead of a damped oscillating pattern is seen in Fig.~\ref{Fig2_large_driving_low_freq} (b) \cite{Sun_PRA_2012,Kenmoe_PRB_2017}. In Fig.~\ref{Fig2_large_driving_low_freq} (c), the left resonator photon number experiences an abrupt change. This is an overall effect of multiple LZ transition interference. In Fig.~\ref{Fig2_large_driving_low_freq} (d) the LZ transition probability of the right qubit tends to be 0.5. Due to a zero energy bias, the right qubit can freely flip with the help of the off-diagonally coupled photons. As a result, there is an equal probability of flipping to the up and down state.

As demonstrated above, we have studied the photon-assisted LZ transitions of the left qubit with the help of the energy diagram. However, due to the strong interference between a large number of LZ transitions, we are not able to analyze each LZ transition explicitly. Therefore, we switch to the case with a strong driving field and a high photon frequency. 

\subsubsection{Qubits coupled to high frequency photons}
\label{Strong driving field and high frequency photons}

Similar to the above case, in this section only the left qubit is driven by a strong periodic field in the absence of the phonon mode, but higher frequency photons with $\omega_{\textrm{r}}=10~\omega_0$ are adopted. When the photon frequency is large, the relaxation time of each LZ transition is much smaller than the time interval between the neighboring LZ transitions. Also, $g<0.1~\omega_{\textrm{r}}$ corresponds to a weak coupling. Therefore, series of independent first-order LZ transitions emerge in the system.

\begin{figure}
  \centering
  \includegraphics[scale=0.5]{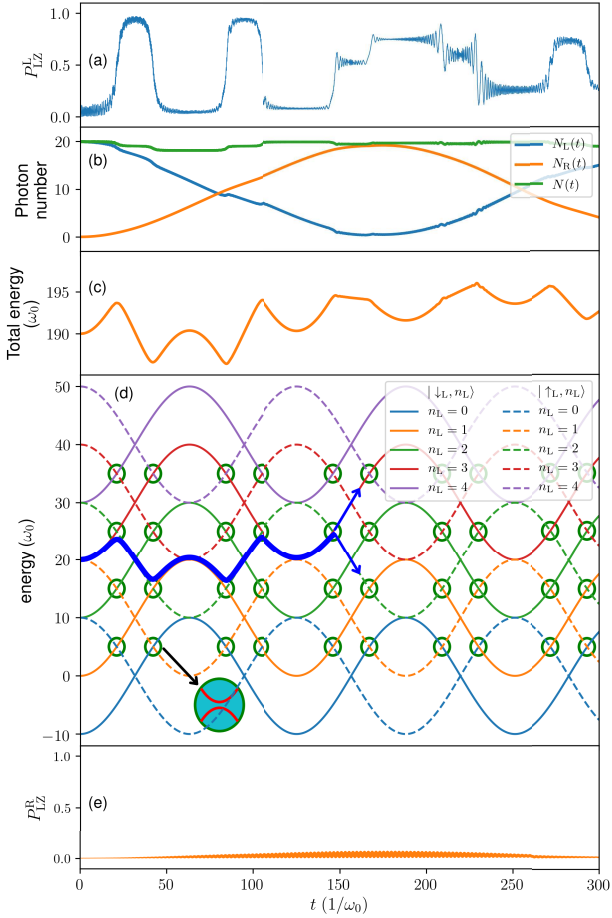} 
  \caption{Dynamics of the Rabi dimer with the qubits coupled to photons with $\omega_{\textrm{r}}=10~\omega_0$, and driven by the fields shown in Fig.~\ref{Fig2_large_driving_low_freq} (a). The LZ transition probabilities of the left and right qubits are illustrated in (a) and (e), respectively. Dynamics of the photon numbers in two resonators is shown in (b). Illustrated in (c) is the dynamics of the total energy of the hybrid system. Panel (d) presents the energy diagram of the coupled states of the left qubit and the number states of the photons in the left resonator. Here we only show the coupled states of the left qubit and the photon number states with $n_{\rm L} = 0,1,...,4$. Since $g<0.1~\omega_{\textrm{r}}$, high order transitions are not allowed. To be specific, $|\downarrow_{\rm L}, n\rangle $ can only transit to $|\uparrow_{\rm L}, n\pm 1 \rangle $, while $|\uparrow_{\rm L}, n\rangle $ can only transit to $|\downarrow_{\rm L}, n\pm 1 \rangle $. Green circles refer to the point where adiabatic transitions may take place. The blue line is the state trajectory if one starts with $|\downarrow_{\rm L}, 3\rangle$}
  \label{Fig3_large_driving_high_freq}
\end{figure}

\begin{figure*}
  \centering
  \includegraphics[scale=0.5]{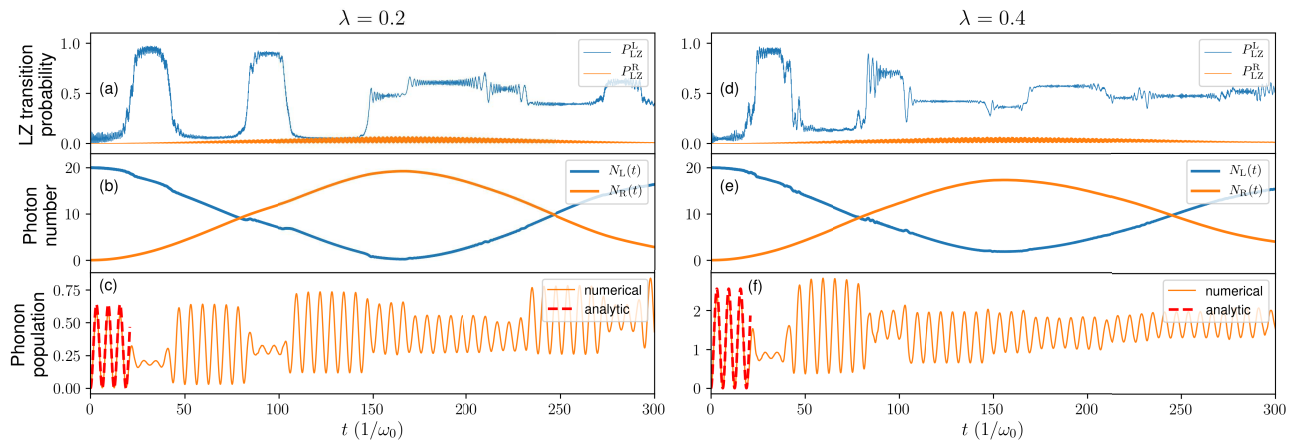}
  \caption{Dynamics of the Rabi dimer with the driven qubits coupled to photons with $\omega_{\textrm{r}}=10~\omega_0$, and a common phonon mode with $\omega_{\textrm{ph}}=\omega_0$. The results for qubit-phonon coupling strengths $\lambda = 0.2$ and 0.4 are presented in the left and right columns, respectively. The upper panels (a) and (d) show the LZ transition probabilities of the qubits. The evolution of the photon numbers are depicted in the middle panels (b) and (e). The bottom panels (c) and (f) exhibit the phonon population dynamics from numerical calculations together with analytic results when $0<t<21/\omega_0$.}
  \label{Fig4_large_driving_high_freq_diss}
\end{figure*}

In Fig.~\ref{Fig3_large_driving_high_freq} (a), the LZ transition probability $P_{\rm LZ}^{\rm L}$ hops to a new value at every transition point as indicated by the green circles. The first four transitions in $0<t<120/\omega_0$ are nearly fully adiabatic transitions due to the large number of off-diagonally coupled photons in the left resonator. As mentioned, a coherent state can be decomposed into an infinite series of number states, and the LZ transition of each number state can be explicitly analyzed with the aid of the energy diagram. In Fig.~\ref{Fig3_large_driving_high_freq} (d), we show a transition trajectory starting with $|\downarrow_{\rm L}, 3 \rangle$ as an example. The first four transitions take place in the sequence of $|\downarrow_{\rm L}, 3 \rangle \rightarrow |\uparrow_{\rm L}, 2\rangle \rightarrow |\downarrow_{\rm L}, 1\rangle \rightarrow |\uparrow_{\rm L}, 2\rangle \rightarrow |\downarrow_{\rm L}, 3\rangle$. We conclude that if every LZ transition is fully adiabatic, starting from any number state $|\downarrow_{\rm L}, n\rangle $ with $n=2,3,4,...$, the state always follows the sequence of $ |\downarrow_{\rm L}, n\rangle \rightarrow |\uparrow_{\rm L}, n-1\rangle \rightarrow |\downarrow_{\rm L}, n-2\rangle \rightarrow |\uparrow_{\rm L}, n-1\rangle \rightarrow |\downarrow_{\rm L}, n\rangle$. Even though $|\downarrow_{\rm L}, 0\rangle$ and $ |\downarrow_{\rm L}, 1\rangle$ cannot initiate the same trajectory, the initial expansion coefficients of these two states are negligibly small for a coherent state $|\downarrow_{\rm L}, \alpha\rangle$ with a large $\alpha$. This can also be explicitly demonstrated by calculating the time evolution of the number state distribution of the photons in the left resonator (cf. left panel of Fig.~\ref{Fig5_photon_distribution_number_states}). In addition to the photon population distribution on the number states, the photon number in the left resonator is crucial to understand the interplay between the photon dynamics and the LZ transitions of the left qubit. As illustrated by the sinusoidal blue curve in Fig.~\ref{Fig3_large_driving_high_freq} (b) in the aforementioned time range, there are four abrupt changes in the photon number in the left resonator with the amplitudes $\Delta N_{\rm L}(t)$ and the time points of the changes summarized in Table~\ref{Left_photon_change}. These changes are more visible in the dynamics of the total photon number $N(t)$ as shown in the green line in Fig.~\ref{Fig3_large_driving_high_freq} (b). These four sudden shifts are attributed to the photon-assisted LZ transitions as we have predicted using the energy diagram. Moreover, during the first four LZ transitions, the system total energy (see Fig.~\ref{Fig3_large_driving_high_freq} (c)) shows precisely the same shape as the path on the energy diagram, which further confirms our analysis.

\begin{table}[htb]
\centering

\begin{tabular}{|c|c|c|c|c|}
\hline
$t(/\omega_0)$ & 21 ($\frac{20\pi}{3}$)  & 42 ($\frac{40\pi}{3}$) & 84 ($\frac{80\pi}{3}$) & 105 ($\frac{100\pi}{3}$) \\
\hline
$\Delta N_{\rm L}(t)$ & -1 & -1 & +1 & +1 \\
\hline
\end{tabular}
\caption{Abrupt changes in the left-resonator photon number induced by photon-assisted LZ transitions.}
\label{Left_photon_change}
\end{table}

At the fifth transition point $t\approx 147/\omega_0$, due to the lack of photons in the left resonator, this is no longer a fully adiabatic transition and the analysis above ceases to work here. Instead, by borrowing the equation of calculating photon-assisted LZ transition probability starting with a coupled state of a left-qubit down-state and a photon coherent state with $N_{\rm L} (t)$ photons in the left resonator~\cite{Sun_PRA_2012}
$$ P_{\rm LZ}^{\rm L}=1-\exp\big[-N_{\rm L}(t)\big] \sum_{n=0}^{\infty} \frac{{N_{\rm L}(t)}^n}{n!}\exp\Big[-\frac{2\pi g^2(n+1)}{v_{\rm L}}\Big],$$
we can get a final probability of LZ transition $P_{\rm LZ}^{\rm L}\approx0.6$ by substituting $N_{\rm L}(t\approx 147/\omega_0)=1$, $g=0.3~\omega_0$, and the sweep rate $v_{\rm L}={\omega_0}^2$ into the equation. The analytical estimation is very close to the probability $P_{\rm LZ}^{\rm L}=0.5$ from numerical calculations as shown in Fig.~\ref{Fig3_large_driving_high_freq} (a) at $t\approx 147/\omega_0$. The minor difference may be attributed to the continuous decreasing of the average photon number in the left resonator during the LZ transition. The fifth transition is shown schematically in Fig.~\ref{Fig3_large_driving_high_freq} (d) in which non-adiabatic and adiabatic transitions occur simultaneously. After the fifth transition, the states will spread out with all photons distributed in a bunch of number states due to the successive LZ transitions, as presented in the left panel of Fig.~\ref{Fig5_photon_distribution_number_states}. Therefore, the transition probabilities are hard to predict analytically afterwards~\cite{Sun_PRA_2012}.

As a reference, the undriven right qubit has a zero bias and is coupled to high-frequency photons with a coupling strength of $g=0.3~\omega_0$. Compared to the photon frequency $\omega_{\textrm{r}}=10~\omega_0$, the off-diagonal qubit-photon coupling is considered to be too weak, $g < 0.1~\omega_{\textrm{r}}$, to alter the state of the right qubit. Moreover, the large detuning between the right qubit and the photons hinders the qubit flipping. Therefore, in contrast to the flipping in the low-frequency photon case with $\omega_{\textrm{r}}=\omega_0$, the right qubit is always confined in its down state when it is coupled to high-frequency photons, as demonstrated in Fig.~\ref{Fig3_large_driving_high_freq} (e).

\subsubsection{Qubits coupled to high frequency photons and a phonon mode}

\begin{figure}
  \centering
  \includegraphics[scale=0.25]{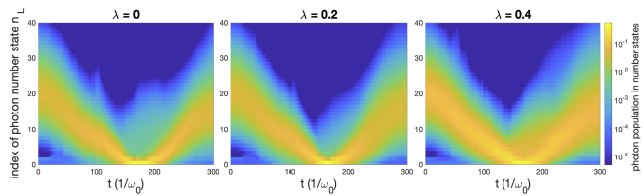}
  \caption{Time evolution of the photon distribution in the number states $|n_L\rangle$ of the left-resonator photons with the qubit-phonon coupling $\lambda = 0$ (left), 0.2 (middle), and 0.4 (right). The $y$ axis represents the index of the number state of the left-resonator photons. Logarithm scaled photon populations are adopted to highlight the photon distribution in the number states under the influence of the phonon mode.}
  \label{Fig5_photon_distribution_number_states}
\end{figure}

In addition to analyzing the LZ transitions in the driven Rabi dimer, we move one step further to study the environmental effects on the LZ transitions by coupling the qubits to a phonon mode. Such mode can be realized in experiments using a micromechanical resonator. In this section, again only the left qubit is driven by a periodic field with $F_{\rm L}=20~\omega_0$, $\Omega_{\rm L}=0.05~\omega_0$, $\Phi_{\rm L}=0$. In addition to the phonon mode with $\omega_{\textrm{ph}}=\omega_0$, two qubits are coupled to high-frequency ($\omega_{\textrm{r}}=10~\omega_0$) photons with $g=0.3~\omega_0$.

In order to highlight the phonon effects on the system dynamics, we perform two sets of calculations with diagonal qubit-phonon couplings $\lambda = 0.2$ and $0.4$. The results are summarized in Figs.~\ref{Fig4_large_driving_high_freq_diss}, in comparison with the results in the absence of the phonon mode as shown in Fig.~\ref{Fig3_large_driving_high_freq}.

Similar to the case without the phonon, the right qubit is still frozen in its down state independent of the qubit-phonon coupling. Therefore, the right photon mode is almost decoupled with the right qubit, leading to the dynamics controlled only by the inter-resonator photon hopping rate $J$ and insensitive to the qubit-phonon coupling. Hereafter, we mainly focus on the analysis of the dynamics of the left qubit and left photon mode, as well as the phonon population.

Comparing Fig.~\ref{Fig3_large_driving_high_freq} (a) with Figs.~\ref{Fig4_large_driving_high_freq_diss} (a) and (d), we find that the left qubit LZ transition probability $P_{\rm LZ}^{\rm L}$ approaches 0.5 faster as $\lambda$ increases, and there are more probability spikes during the LZ transitions. The underlying reasons are shown below.

Coupling the qubits to low-frequency phonon induces more split states with smaller energy gaps. Therefore, more transitions between these states are possible to take place, leading to fluctuations in the LZ transition probability of the left qubit. More transitions also accelerate the wave packet spreading over more states of the system, and thus results in equally populated up and down states of the left qubit in a shorter time.

To further verify our explanation, we record the real-time dynamics of the left-resonator photon distribution in the number states with different qubit-phonon coupling strengths (Fig.~\ref{Fig5_photon_distribution_number_states}). From a general point of view, the number states with the highest probability always have a photon number around the real-time average photon number $N_{\rm L}(t)$ in the left resonator, which is the result of the Poissonian distribution. However, as the qubit-phonon coupling strength $\lambda$ increases, the photon population spreads to a broader range of number states which implies that the states branch out and cause $P_{\rm LZ}^{\rm L}\rightarrow 0.5$. In a conventional, linearly driven LZ model, the zero-temperature dissipative environment is not able to change the LZ transition probability \cite{Saito_PRB_2007}. However, our model is much more intricate with a sinusoidal driving field and tunneling photons. It has been shown that even at zero temperature, sinusoidal driving is able to change the final LZ transition probability. Also, stronger the coupling strength, larger the LZ transition probability deviation \cite{Ota_APE_2017}. Moreover, it has been found that a large coupling strength $\lambda$ and a small initial phonon number enable the system to enter a fully quantum regime which cannot be predicted by the semi-classical model. In this regime, the transition probability decays faster to 0.5 with sharp peaks of a few frequency components~\cite{Ashhab_JPA_2017}. This is consistent with our findings shown in Figs.~\ref{Fig4_large_driving_high_freq_diss} (a) and (d).

It is an intriguing fact that the phonon population oscillates sinusoidally with a fixed frequency but varying amplitudes dependent on the state of the left qubit. Moreover, the left qubit in the up state can greatly damp the oscillations while the down state facilitates the oscillations. We can approach this problem by analyzing the photon population dynamics before the first LZ transition, i.e., $t<21/\omega_0$. In this period, even with a large number of off-diagonally coupled photons in the left resonator, the qubit state is confined in its initial down state due to the large energy gap between the up and the down states. Therefore, we can neglect the contribution of the off-diagonally coupled photons to the phonon population, i.e., only consider the two qubits and diagonally coupled phonon. This reasonable approximation enables us to get an analytical solution (shown in Appendix \ref{phonon_population}) of the phonon population dynamics due to the temporary time independence of both $\sigma^{\rm L}_z$ and $\sigma^{\rm R}_z$ operators in the Heisenberg space. The analytic result $N_{\textrm{ph}}^{\rm A} (t)=8\lambda^2\big[1-\cos(\omega_{\textrm{ph}} t)\big]/\omega_{\textrm{ph}}^2$ when $t<21/\omega_0$ is consistent with the results obtained by our numerical calculation as shown in Figs.~\ref{Fig4_large_driving_high_freq_diss} (c) and (f). The minor difference is induced by our approximation of neglecting the contribution from the photons.

In the time interval [$21/\omega_0$, $42/\omega_0$], the left qubit flips to the up state and the oscillating amplitude of the phonon population is suppressed. This can be explained by the expression of the number operator $b^{\dagger}_{(t)}b_{(t)}$ in the Heisenberg picture, which contains the term $\sigma^{\rm L}_z+\sigma^{\rm R}_z$. The right qubit is always suppressed in its down state and hence if the left qubit is in the up state, the coefficients of the time dependent oscillating terms are zero. The small amplitude oscillation is due to the fact that the left qubit is not in an exact up state.
 
Even though the photons start to play an important role in the system afterwards, the oscillation amplitude of the phonon population is still dependent on the left qubit state and the oscillation frequency is always $\omega_{\textrm{ph}}$. This is due to the lack of direct interactions between the photons and the phonon. However, we see a slow increase in the phonon population. The phonon mode gradually gains energy from photons through the intermediate qubit.

\subsection{Qubit and photon dynamics manipulation by driving both qubits}

\begin{figure}
  \centering
  \includegraphics[scale=0.55]{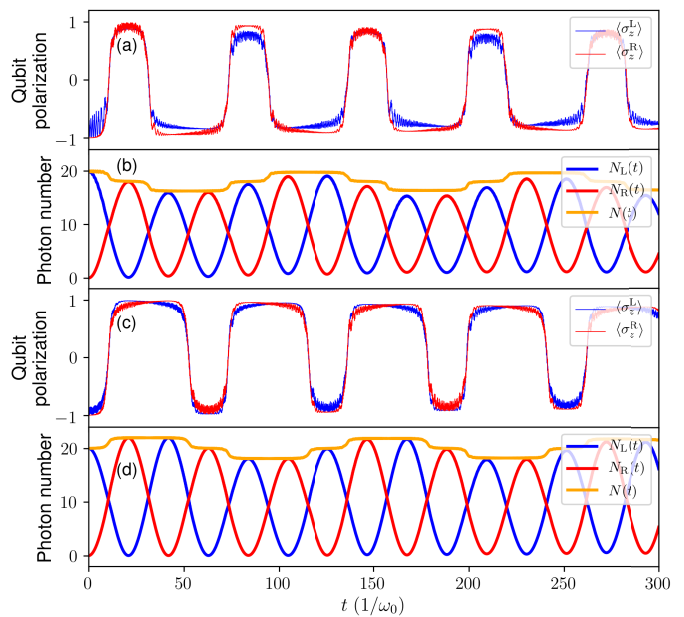}
  \caption{ (a)-(b) Qubit polarization and photon dynamics with $\Phi_{\rm L} = \Phi_{\rm R} = \pi/6$. (c)-(d) Qubit polarization and photon dynamics with $\Phi_{\rm L} = \Phi_{\rm R} = \pi/2$. The other parameters for the fields are $F_{\rm L} = F_{\rm R} = 20~\omega_0$, and $\Omega_{\rm L} = \Omega_{\rm R} =0.05~\omega_0$. High frequency photons with $\omega_{\textrm{r}}=10~\omega_0$ can hop between two resonators with $J=0.075~\omega_0$. The qubits are not coupled to the phonon mode, i.e., $\lambda=0$. }
  \label{Fig6_manipulate}
\end{figure}

Manipulating the states of the qubits and the photons is one of the main tasks in quantum information science and quantum computing \cite{Gao_Nature_Photonics_2015}. For example, Ku {\it et al.} applied a microwave hyperbolic secant pulse to tune a superconducting transmon qubit and generated high-fidelity phase gates \cite{Ku_PRA_2017}. We have also proposed a framework to engineer the photon dynamics in a Rabi dimer by tunning the qubit-phonon coupling \cite{ZFL_AdP_2018}. In this work, we find that the dynamics of the qubits and photons in the Rabi dimer can be precisely manipulated by tunning the external driving fields and the inter-resonator photon tunneling rate.

Here we propose a solution of generating a square-wave pattern in qubit dynamics together with a stepwise photon dynamics. To illustrate our scheme, we parameterize the system as follows: $J=0.075~\omega_0$, $F_{\rm L} = F_{\rm R} = 20~\omega_0$, $\Omega_{\rm L} = \Omega_{\rm R} =0.05~\omega_0$, $\omega_{\textrm{r}}=10~\omega_0$, $\lambda=0$. With a large photon hopping rate, i.e., $J=0.075~\omega_0$, a relatively large number of photons can reside in each resonator at the transition times. The large photon numbers lead to LZ transitions with almost hundred percent probability, which in turn prevent uncontrollable spreading out of the states.

Two scenarios which differ only in the initial phase of the external field are recommended here. It is found that the initial phase of the external driving field plays an important role in controlling both the qubit and photon dynamics starting from the same initial state. As shown in Fig.~\ref{Fig6_manipulate}, the polarization $\langle \sigma_z \rangle$ of both qubits exhibits the square-wave patterns. With external fields applied to both qubits, simultaneous LZ transitions in two qubits lead to a larger photon number variation in the system. Whenever there are qubit state transitions, the photon number will either increase $2$ or decrease $2$. When $\Phi_{\rm L} =\Phi_{\rm R}=\pi/6$, both qubits are in their down states most of the time while in the up states with $\Phi_{\rm L} =\Phi_{\rm R}=\pi/2$. In addition, the photon number can vary in a range of $16-20$ for $\Phi_{\rm L} =\Phi_{\rm R}=\pi/6$ while $18-22$ for $\Phi_{\rm L} =\Phi_{\rm R}=\pi/2$.

All the phenomena mentioned in this subsection can be well explained by the energy diagram shown in section \ref{Strong driving field and high frequency photons} and hence we do not show the diagram again. As illustrated by the examples in this section, the states of the qubits and the photons can be engineered by tuning the driving fields on the qubits. It opens up new venues to control quantum states in quantum information devices and quantum computers.

\section{Conclusion}\label{conclusions}

In this work, we have found intriguing multiple photon-assisted LZ transitions in a Rabi dimer system with the qubits driven by periodic fields. Combining the multiple Davydov D$_2$ {\it Ansatz} with the Dirac-Frenkel time-dependent variational principle, we investigate the dynamics of the qubits, the photon modes, and the phonon mode explicitly. Driving the left qubit with a periodic field, we record the time-dependent LZ transition probability of the qubit. Both adiabatic and nonadiabatic transitions are involved in the dynamics, which is well explained by the energy digram and the system energy. The trajectory for the transitions is also revealed using the energy diagram. If the qubits are further coupled to a common phonon mode, the LZ transition probability approach 0.5 faster as the phonon mode introduces a large number of available states with small energy gaps. The short time phonon population is estimated analytically, which agrees well with numerical calculations.

Equipped the knowledge of controlling the LZ transitions, we propose methods of manipulating the qubit states and the photon dynamics with the help of photon-assisted LZ transitions. The circuit QED setup proposed in this work could also be used to prepare quantum states using the Landau-Zener-Stückelberg interferometry \cite{Shevchenko_PR_2010, Lidal_PRA_2020, Bonifacio_PRB_2020}. Studies along this direction are in progress. It is expected that these scenarios can be experimentally realized and may benefit the quantum information technology.

\section{Acknowledgement}
The authors gratefully acknowledge the support of the Singapore Ministry of Education Academic Research Fund (Grant Nos.~2018-T1-002-175 and 2020-T1-002-075). YJS would like to acknowledge the funding support from Nanyang Technological University – URECA Undergraduate Research Programme. This research was also supported in part by Zhejiang Provincial Natural Science Foundation of China under Grant No.~LY18A040005.

\section*{Data Availability}

The data that support the findings of this study are available from the corresponding author upon reasonable request.

\appendix
\section{The time dependent variational approach}
\label{Equations of Motion}

The Dirac-Frenkel variational principle results in equations of motion of the variational parameters as follows,
\begin{eqnarray}  
&&i\sum_{n=1}^{M}\left[\dot{A}_{n}+A_{n}\left(\mu_{l}^{*}\dot{\mu}_{n}+\nu_{l}^{*}\dot{\nu}_{n}+\eta_{n}^{*}\dot{\eta}_{n}\right)\right]\bar{S}_{ln}\nonumber\\
&&=\sum_{n=1}^{M}A_{n}\bar{S}_{ln}\left[\frac{F_{\rm L}}{2}\cos\left(\Omega_{\rm L}t+\text{\ensuremath{\Phi_{\rm L}}}\right)+\frac{F_{\rm R}}{2}\cos\left(\Omega_{\rm R}t+\text{\ensuremath{\Phi_{\rm R}}}\right)\right.\nonumber\\
&&+\omega_{\rm L}\text{\ensuremath{\mu_{l}^{*}}}\mu_{n}+\omega_{\rm R}\text{\ensuremath{\nu_{l}^{*}}}\nu_{n}-J\left(\mu_{l}^{*}\nu_{n}+\mu_{n}\nu_{l}^{*}\right)\nonumber\\
&&\left.+\omega_{\textrm{ph}}\eta_{l}^{*}\eta_{n}+2\lambda\left(\eta_{l}^{*}+\eta_{n}\right)\right]\nonumber\\
&&-g\sum_{n=1}^{M}C_{n}\left(\mu_{l}^{*}+\mu_{n}\right)\bar{S}_{ln}-g\sum_{n=1}^{M}B_{n}\left(\nu_{l}^{*}+\nu_{n}\right)\bar{S}_{ln},
\end{eqnarray}
where $\bar{S}_{ln}=\exp\left[\mu_{l}^{*}\mu_{n}+\nu_{l}^{*}\nu_{n}+\eta_{l}^{*}\eta_{n}\right]$.

\begin{eqnarray}   
&&i\sum_{n=1}^{M}\left[\dot{B}_{n}+B_{n}\left(\mu_{l}^{*}\dot{\mu}_{n}+\nu_{l}^{*}\dot{\nu}_{n}+\eta_{n}^{*}\dot{\eta}_{n}\right)\right]\bar{S}_{ln}\nonumber\\
&&=  \sum_{n=1}^{M}B_{n}\bar{S}_{ln}\left[\frac{F_{\rm L}}{2}\cos\left(\Omega_{\rm L}t+\text{\ensuremath{\Phi_{\rm L}}}\right)-\frac{F_{\rm R}}{2}\cos\left(\Omega_{\rm R}t+\text{\ensuremath{\Phi_{\rm R}}}\right)\right.\nonumber\\
&&\left.+\omega_{\rm L}\text{\ensuremath{\mu_{l}^{*}}}\mu_{n}+\omega_{\rm R}\text{\ensuremath{\nu_{l}^{*}}}\nu_{n}-J\left(\mu_{l}^{*}\nu_{n}+\mu_{n}\nu_{l}^{*}\right)+\omega_{\textrm{ph}}\eta_{l}^{*}\eta_{n}\right]\nonumber\\
&&-g\sum_{n=1}^{M}D_{n}\left(\mu_{l}^{*}+\mu_{n}\right)\bar{S}_{ln}-g\sum_{n=1}^{M}A_{n}\left(\nu_{l}^{*}+\nu_{n}\right)\bar{S}_{ln},
\end{eqnarray}

\begin{eqnarray}   &&i\sum_{n=1}^{M}\left[\dot{C}_{n}+C_{n}\left(\mu_{l}^{*}\dot{\mu}_{n}+\nu_{l}^{*}\dot{\nu}_{n}+\eta_{n}^{*}\dot{\eta}_{n}\right)\right]\bar{S}_{ln}\nonumber\\
&&=  \sum_{n=1}^{M}C_{n}\bar{S}_{ln}\left[-\frac{F_{\rm L}}{2}\cos\left(\Omega_{\rm L}t+\text{\ensuremath{\Phi_{\rm L}}}\right)+\frac{F_{\rm R}}{2}\cos\left(\Omega_{\rm R}t+\text{\ensuremath{\Phi_{\rm R}}}\right)\right.\nonumber\\
&&\left.+\omega_{\rm L}\text{\ensuremath{\mu_{l}^{*}}}\mu_{n}+\omega_{\rm R}\text{\ensuremath{\nu_{l}^{*}}}\nu_{n}-J\left(\mu_{l}^{*}\nu_{n}+\mu_{n}\nu_{l}^{*}\right)+\omega_{\textrm{ph}}\eta_{l}^{*}\eta_{n}\right]\nonumber\\
&&-g\sum_{n=1}^{M}A_{n}\left(\mu_{l}^{*}+\mu_{n}\right)\bar{S}_{ln}-g\sum_{n=1}^{M}D_{n}\left(\nu_{l}^{*}+\nu_{n}\right)\bar{S}_{ln},
\end{eqnarray}

\begin{eqnarray}   
&&i\sum_{n=1}^{M}\left[\dot{D}_{n}+D_{n}\left(\mu_{l}^{*}\dot{\mu}_{n}+\nu_{l}^{*}\dot{\nu}_{n}+\eta_{n}^{*}\dot{\eta}_{n}\right)\right]\bar{S}_{ln}\nonumber\\
&&=  \sum_{n=1}^{M}D_{n}\bar{S}_{ln}\left[-\frac{F_{\rm L}}{2}\cos\left(\Omega_{\rm L}t+\text{\ensuremath{\Phi_{\rm L}}}\right)-\frac{F_{\rm R}}{2}\cos\left(\Omega_{\rm R}t+\text{\ensuremath{\Phi_{\rm R}}}\right)\right.\nonumber\\
&&+\omega_{\rm L}\text{\ensuremath{\mu_{l}^{*}}}\mu_{n}+\omega_{\rm R}\text{\ensuremath{\nu_{l}^{*}}}\nu_{n}-J\left(\mu_{l}^{*}\nu_{n}+\mu_{n}\nu_{l}^{*}\right)\nonumber\\
&&\left.+\omega_{\textrm{ph}}\eta_{l}^{*}\eta_{n}-2\lambda\left(\eta_{l}^{*}+\eta_{n}\right)\right]\nonumber\\
&&-g\sum_{n=1}^{M}B_{n}\left(\mu_{l}^{*}+\mu_{n}\right)\bar{S}_{ln}-g\sum_{n=1}^{M}C_{n}\left(\nu_{l}^{*}+\nu_{n}\right)\bar{S}_{ln},
\end{eqnarray}

\begin{align}
\begin{autobreak}
\MoveEqLeft
i\sum_{n=1}^{M} \left[\left(A_{l}^{*}\dot{A}_{n}+B_{l}^{*}\dot{B}_{n}+C_{l}^{*}\dot{C_{n}}+D_{l}^{*}\dot{D}_{n}\right)\mu_{n}\right.
+\left(A_{l}^{*}A_{n}+B_{l}^{*}B_{n}+C_{l}^{*}C_{n}+D_{l}^{*}D_{n}\right)\dot{\mu}{}_{n}
+\left(A_{l}^{*}A_{n}+B_{l}^{*}B_{n}+C_{l}^{*}C_{n}+D_{l}^{*}D_{n}\right)
\left.\times\left(\mu_{l}^{*}\dot{\mu}_{n}+\nu_{l}^{*}\dot{\nu}_{n}+\eta_{n}^{*}\dot{\eta}_{n}\right)\mu_{n}\right]\bar{S}_{ln}
=i\sum_{n=1}^{M}\frac{\mu_{n}}{2}\left[\left(A_{l}^{*}A_{n}+B_{l}^{*}B_{n}-C_{l}^{*}C_{n}-D_{l}^{*}D_{n}\right)\right.
\left.\times F_{\rm L}\cos\left(\omega_{\rm L}t+\text{\ensuremath{\Phi_{\rm L}}}\right)\right]\bar{S}_{ln}
+i\sum_{n=1}^{M}\frac{\mu_{n}}{2}\left[\left(A_{l}^{*}A_{n}-B_{l}^{*}B_{n}+C_{l}^{*}C_{n}-D_{l}^{*}D_{n}\right)\right.
\left.\times F_{\rm R} \cos\left(\omega_{\rm R}t+\text{\ensuremath{\Phi_{\rm R}}}\right)\right]\bar{S}_{ln}
+\sum_{n=1}^{M}\mu_{n}\bar{S}_{ln}\left(A_{l}^{*}A_{n}+B_{l}^{*}B_{n}+C_{l}^{*}C_{n}+D_{l}^{*}D_{n}\right)
\times[\omega_{\rm L}\text{\ensuremath{\mu_{l}^{*}}}\mu_{n}+\omega_{\rm R}\text{\ensuremath{\nu_{l}^{*}}}\nu_{n}-J\left(\mu_{l}^{*}\nu_{n}+\mu_{n}\nu_{l}^{*}\right)+\omega_{\textrm{ph}}\eta_{l}^{*}\eta_{n}]
+\sum_{n=1}^{M}\bar{S}_{ln}\left(A_{l}^{*}A_{n}+B_{l}^{*}B_{n}+C_{l}^{*}C_{n}+D_{l}^{*}D_{n}\right)
\times\left(\omega_{\rm L}\mu_{n}-J\nu_{n}\right)
-g\sum_{n=1}^{M}\mu_{n}\left(A_{l}^{*}C_{n}+B_{l}^{*}D_{n}+C_{l}^{*}A_{n}+D_{l}^{*}B_{n}\right)
\times\left(\mu_{l}^{*}+\mu_{n}\right)\bar{S}_{ln}
-g\sum_{n=1}^{M}\mu_{n}\left(A_{l}^{*}C_{n}+B_{l}^{*}D_{n}+C_{l}^{*}A_{n}+D_{l}^{*}B_{n}\right)\bar{S}_{ln}
-g\sum_{n=1}^{M}\mu_{n}\left(A_{l}^{*}B_{n}+B_{l}^{*}A_{n}+C_{l}^{*}D_{n}+D_{l}^{*}C_{n}\right)
\times\left(\nu_{l}^{*}+\nu_{n}\right)\bar{S}_{ln}
+\sum_{n=1}^{M}\mu_{n}\left(2A_{l}^{*}A_{n}-D_{l}^{*}D_{n}\right)\lambda\left(\eta_{l}^{*}+\eta_{n}\right)\bar{S}_{ln},
\end{autobreak}
\end{align}

\begin{align}
\begin{autobreak}
\MoveEqLeft
i\sum_{n=1}^{M} \left[\left(A_{l}^{*}\dot{A}_{n}+B_{l}^{*}\dot{B}_{n}+C_{l}^{*}\dot{C_{n}}+D_{l}^{*}\dot{D}_{n}\right)\nu_{n}\right.
+\left(A_{l}^{*}A_{n}+B_{l}^{*}B_{n}+C_{l}^{*}C_{n}+D_{l}^{*}D_{n}\right)\dot{\nu}{}_{n}
+\left(A_{l}^{*}A_{n}+B_{l}^{*}B_{n}+C_{l}^{*}C_{n}+D_{l}^{*}D_{n}\right)
\left.\times\left(\mu_{l}^{*}\dot{\mu}_{n}+\nu_{l}^{*}\dot{\nu}_{n}+\eta_{n}^{*}\dot{\eta}_{n}\right)\nu_{n}\right]\bar{S}_{ln}
= i\sum_{n=1}^{M}\frac{\nu_{n}}{2}\left[\left(A_{l}^{*}A_{n}+B_{l}^{*}B_{n}-C_{l}^{*}C_{n}-D_{l}^{*}D_{n}\right)\right.
\left.\times F_{\rm L} \cos\left(\omega_{\rm L}t+\text{\ensuremath{\Phi_{\rm L}}}\right)\right]\bar{S}_{ln}
+ i\sum_{n=1}^{M}\frac{\nu_{n}}{2}\left[\left(A_{l}^{*}A_{n}-B_{l}^{*}B_{n}+C_{l}^{*}C_{n}-D_{l}^{*}D_{n}\right)\right.
\left.\times F_{\rm R} \cos\left(\omega_{\rm R}t+\text{\ensuremath{\Phi_{\rm R}}}\right)\right]\bar{S}_{ln}
+\sum_{n=1}^{M}\nu_{n}\bar{S}_{ln}\left(A_{l}^{*}A_{n}+B_{l}^{*}B_{n}+C_{l}^{*}C_{n}+D_{l}^{*}D_{n}\right)
\times[\omega_{\rm L}\text{\ensuremath{\mu_{l}^{*}}}\mu_{n}+\omega_{\rm R}\text{\ensuremath{\nu_{l}^{*}}}\nu_{n}-J\left(\mu_{l}^{*}\nu_{n}+\mu_{n}\nu_{l}^{*}\right)+\omega_{\textrm{ph}}\eta_{l}^{*}\eta_{n}]
+\sum_{n=1}^{M}\bar{S}_{ln}\left(A_{l}^{*}A_{n}+B_{l}^{*}B_{n}+C_{l}^{*}C_{n}+D_{l}^{*}D_{n}\right)
\times\left(\omega_{\rm R}\nu_{n}-J\mu_{n}\right)
-g\sum_{n=1}^{M}\nu_{n}\left(A_{l}^{*}C_{n}+B_{l}^{*}D_{n}+C_{l}^{*}A_{n}+D_{l}^{*}B_{n}\right)
\times\left(\mu_{l}^{*}+\mu_{n}\right)\bar{S}_{ln}
-g\sum_{n=1}^{M}\nu_{n}\left(A_{l}^{*}B_{n}+B_{l}^{*}A_{n}+C_{l}^{*}D_{n}+D_{l}^{*}C_{n}\right)\bar{S}_{ln}
-g\sum_{n=1}^{M}\nu_{n}\left(A_{l}^{*}B_{n}+B_{l}^{*}A_{n}+C_{l}^{*}D_{n}+D_{l}^{*}C_{n}\right)
\times\left(\nu_{l}^{*}+\nu_{n}\right)\bar{S}_{ln} 
+\sum_{n=1}^{M}\nu_{n}\left(2A_{l}^{*}A_{n}-D_{l}^{*}D_{n}\right)\lambda\left(\eta_{l}^{*}+\eta_{n}\right)\bar{S}_{ln},
\end{autobreak}
\end{align}
and
\begin{align}
\begin{autobreak}
\MoveEqLeft 
i\sum_{n=1}^{M}\left[\left(A_{l}^{*}\dot{A}_{n}+B_{l}^{*}\dot{B}_{n}+C_{l}^{*}\dot{C_{n}}+D_{l}^{*}\dot{D}_{n}\right)\eta_{n}\right.
+\left(A_{l}^{*}A_{n}+B_{l}^{*}B_{n}+C_{l}^{*}C_{n}+D_{l}^{*}D_{n}\right)\dot{\eta}{}_{n}
+\left(A_{l}^{*}A_{n}+B_{l}^{*}B_{n}+C_{l}^{*}C_{n}+D_{l}^{*}D_{n}\right)
\left.\times\left(\mu_{l}^{*}\dot{\mu}_{n}+\nu_{l}^{*}\dot{\nu}_{n}+\eta_{n}^{*}\dot{\eta}_{n}\right)\eta_{n}\right]\bar{S}_{ln}
= 
i\sum_{n=1}^{M}\frac{\eta_{n}}{2}\left[\left(A_{l}^{*}A_{n}+B_{l}^{*}B_{n}-C_{l}^{*}C_{n}-D_{l}^{*}D_{n}\right)\right. 
\left.\times F_{\rm L}\cos\left(\omega_{\rm L}t+\text{\ensuremath{\Phi_{\rm L}}}\right)\right]\bar{S}_{ln}
+i\sum_{n=1}^{M}\frac{\eta_{n}}{2}\left[\left(A_{l}^{*}A_{n}-B_{l}^{*}B_{n}+C_{l}^{*}C_{n}-D_{l}^{*}D_{n}\right)\right.  
\left.\times F_{\rm R} \cos\left(\omega_{\rm R}t+\text{\ensuremath{\Phi_{\rm R}}}\right)\right]\bar{S}_{ln}
+\sum_{n=1}^{M}\eta_{n}\bar{S}_{ln}\left(A_{l}^{*}A_{n}+B_{l}^{*}B_{n}+C_{l}^{*}C_{n}+D_{l}^{*}D_{n}\right)
\times[\omega_{\rm L}\text{\ensuremath{\mu_{l}^{*}}}\mu_{n}+\omega_{\rm R}\text{\ensuremath{\nu_{l}^{*}}}\nu_{n}-J\left(\mu_{l}^{*}\nu_{n}+\mu_{n}\nu_{l}^{*}\right)+\omega_{\textrm{ph}}\eta_{l}^{*}\eta_{n}]
+\sum_{n=1}^{M}\left(A_{l}^{*}A_{n}+B_{l}^{*}B_{n}+C_{l}^{*}C_{n}+D_{l}^{*}D_{n}\right)\omega_{\textrm{ph}}\eta_{n}\bar{S}_{ln}
-g\sum_{n=1}^{M}\eta_{n}\left(A_{l}^{*}C_{n}+B_{l}^{*}D_{n}+C_{l}^{*}A_{n}+D_{l}^{*}B_{n}\right)
\times\left(\mu_{l}^{*}+\mu_{n}\right)\bar{S}_{ln}  
-g\sum_{n=1}^{M}\eta_{n}\left(A_{l}^{*}B_{n}+B_{l}^{*}A_{n}+C_{l}^{*}D_{n}+D_{l}^{*}C_{n}\right)
\times\left(\nu_{l}^{*}+\nu_{n}\right)\bar{S}_{ln}
+\sum_{n=1}^{M}\eta_{n}\left(2A_{l}^{*}A_{n}-2D_{l}^{*}D_{n}\right)\lambda\left(\eta_{l}^{*}+\eta_{n}\right)\bar{S}_{ln}
+\sum_{n=1}^{M}\left(2A_{l}^{*}A_{n}-D_{l}^{*}D_{n}\right)\lambda\bar{S}_{ln}.
\end{autobreak}
\end{align}

By numerically solving these linear equations at each time $t$, one can calculate the values of $\dot{A}_{n}$, $\dot{B}_{n}$, $\dot{C}_{n}$, $\dot{D}_{n}$, $\dot{\mu}_{n}$, $\dot{\nu}_{n}$, and $\dot{\eta}_{n}$ accurately. The fourth-order Runge-Kutta method is then adopted for the time evolution of the tunable Rabi dimer, including the time-dependent photon numbers, phonon number, qubit polarization, LZ transition probability.

\section{Approximated analytical solution of phonon population}
\label{phonon_population}

The Hamiltonian of the simplified system is
\begin{eqnarray}
&& H_{\rm s}=\frac{F_{\rm L} \cos(\Omega_{\rm L}t+\Phi_{\rm L})}{2}\sigma^{\rm L}_z+\frac{F_{\rm R} \cos(\Omega_{\rm R}t+\Phi_{\rm R})}{2}\sigma^{\rm R}_z\nonumber\\
&& ~~~~+ \omega_{\textrm{ph}} b^{\dagger}b+\lambda(\sigma^{\rm L}_z+\sigma^{\rm R}_z)(b^{\dagger}+b) \nonumber
\end{eqnarray}

Now we use Heisenberg picture to calculate the dynamics of phonon number which start at  vacuum state $|0 \rangle$. We use Heisenberg equation of motion:
$$\frac{dQ^{(H_{\rm s})}}{dt}=\frac{1}{i\hbar}[Q^{(H_{\rm s})},H_{\rm s}],$$ where $Q$ is any operator. Here we set $\hbar=1$ and therefore the equation becomes $$\frac{dQ^{(H_{\rm s})}}{dt}=\frac{1}{i}[Q^{(H_{\rm s})},H_{\rm s}]$$\\
By solving this differential equation we can obtain the time evolution of the  operator in Heisenberg space and thus calculate the real-time average phonon number in the system by the equation$$N_{\textrm{ph}} (t)=\langle\Phi_{(0)}|b^{\dagger}_{(t)}b_{(t)}|\Phi_{(0)} \rangle$$ where $| \Phi_{(0)} \rangle = |\downarrow^{\rm L}, \downarrow^{\rm R},0 \rangle$ is the initial state of two qubits and phonon.\\

we next calculate the operator $b^{\dagger}_{(t)}$ and $b_{(t)}$ in Heisenberg space respectively.\\

\begin{equation}
\begin{split}
\frac{db^{\dagger}}{dt} & = \frac{1}{i}[b^{\dagger},H_{\rm s}] \\
 & = \frac{1}{i}[b^{\dagger},\frac{F_{\rm L} \cos(\Omega_{\rm L}t+\Phi_{\rm L})}{2}\sigma^{\rm L}_z+\frac{F_{\rm R} \cos(\Omega_{\rm R}t+\Phi_{\rm R})}{2}\sigma^{\rm R}_z\\
 &~~~+\omega_{\textrm{ph}} b^{\dagger}b+\lambda(\sigma^{\rm L}_z+\sigma^{\rm R}_z)(b^{\dagger}+b)] \\
 &= \frac{1}{i}[b^{\dagger},\omega_{\textrm{ph}} b^{\dagger}b+\lambda(\sigma^{\rm L}_z+\sigma^{\rm R}_z)(b^{\dagger}+b)] \\
  &=\frac{\omega_{\textrm{ph}}}{i}[b^{\dagger},b^{\dagger}b]+\frac{\lambda(\sigma^{\rm L}_z+\sigma^{\rm R}_z)}{i}[b^{\dagger},b^{\dagger}+b] \\
  &=-\frac{\omega_{\textrm{ph}}}{i}b^{\dagger}-\frac{\lambda(\sigma^{\rm L}_z+\sigma^{\rm R}_z)}{i} \\
  &=i\omega_{\textrm{ph}} b^{\dagger}+i\lambda(\sigma^{\rm L}_z+\sigma^{\rm R}_z)
\end{split}
\end{equation}

\begin{equation}
\begin{split}
\frac{db}{dt} & = \frac{1}{i}[b,H_{\rm s}] \\
 & = \frac{1}{i}[b,\frac{F_{\rm L} \cos(\Omega_{\rm L}t+\Phi_{\rm L})}{2}\sigma^{\rm L}_z+\frac{F_{\rm R} \cos(\Omega_{\rm R}t+\Phi_{\rm R})}{2}\sigma^{\rm R}_z\\
 &~~~+\omega_{\textrm{ph}} b^{\dagger}b+\lambda(\sigma^{\rm L}_z+\sigma^{\rm R}_z)(b^{\dagger}+b)] \\
 &= \frac{1}{i}[b,\omega_{\textrm{ph}} b^{\dagger}b+\lambda(\sigma^{\rm L}_z+\sigma^{\rm R}_z)(b^{\dagger}+b)] \\
  &=\frac{\omega_{\textrm{ph}}}{i}[b,b^{\dagger}b]+\frac{\lambda(\sigma^{\rm L}_z+\sigma^{\rm R}_z)}{i}[b,b^{\dagger}+b] \\
  &=\frac{\omega_{\textrm{ph}}}{i}b+\frac{\lambda(\sigma^{\rm L}_z+\sigma^{\rm R}_z)}{i} \\
  &=-i\omega_{\textrm{ph}} b-i\lambda(\sigma^{\rm L}_z+\sigma^{\rm R}_z)
\end{split}
\end{equation}

where we have used the fact that $[b,b^{\dagger}]=\mathbb{I}$

Next,we show that $\sigma^{\rm L}_z$ and $\sigma^{\rm R}_z$ are time independent which enable us to solve the above two differential equations.

\begin{equation}
\begin{split}
\frac{d\sigma^{\rm L}_z}{dt} & = \frac{1}{i}[\sigma^{\rm L}_z,H_{\rm s}] \\
 & = \frac{1}{i}[\sigma^{\rm L}_z,\frac{F_{\rm L} \cos(\Omega_{\rm L}t+\Phi_{\rm L})}{2}\sigma^{\rm L}_z+\frac{F_{\rm R} \cos(\Omega_{\rm R}t+\Phi_{\rm R})}{2}\sigma^{\rm R}_z\\
 &~~~+\omega_{\textrm{ph}} b^{\dagger}b+\lambda(\sigma^{\rm L}_z+\sigma^{\rm R}_z)(b^{\dagger}+b)] \\
  & = \frac{1}{i}[\sigma^{\rm L}_z,\frac{F_{\rm L} \cos(\Omega_{\rm L}t+\Phi_{\rm L})}{2}\sigma^{\rm L}_z+\lambda\sigma^{\rm L}_z(b^{\dagger}+b)]\\
  &=0
\end{split}
\end{equation}

\begin{equation}
\begin{split}
\frac{d\sigma^{\rm R}_z}{dt} & = \frac{1}{i}[\sigma^{\rm R}_z,H_{\rm s}] \\
 & = \frac{1}{i}[\sigma^{\rm R}_z,\frac{F_{\rm L} \cos(\Omega_{\rm L}t+\Phi_{\rm L})}{2}\sigma^{\rm L}_z+\frac{F_{\rm R} \cos(\Omega_{\rm R}t+\Phi_{\rm R})}{2}\sigma^{\rm R}_z\\
 &~~~+\omega_{\textrm{ph}} b^{\dagger}b+\lambda(\sigma^{\rm L}_z+\sigma^{\rm R}_z)(b^{\dagger}+b)] \\
 & = \frac{1}{i}[\sigma^{\rm R}_z,\frac{F_{\rm R} \cos(\Omega_{\rm R}t+\Phi_{\rm R})}{2}\sigma^{\rm R}_z+\lambda\sigma^{\rm R}_z(b^{\dagger}+b)] \\
  &=0
\end{split}
\end{equation}

Now we solve the differential equation. we choose $b^{\dagger}$ as an example and will provide the solution for $b$ directly.

\begin{equation}
\begin{split}
&\frac{db^{\dagger}}{dt}=i\omega_{\textrm{ph}} b^{\dagger}+i\lambda(\sigma^{\rm L}_z+\sigma^{\rm R}_z) \\
&\int_{b^{\dagger}_{(0)}}^{b^{\dagger}_{(t)}} \frac{db^{\dagger}}{i\omega_{\textrm{ph}} b^{\dagger}+i\lambda(\sigma^{\rm L}_z+\sigma^{\rm R}_z)}=\int_{0}^{t} dt\\
&b^{\dagger}_{(t)}=\frac{e^{i\omega_{\textrm{ph}} t}\big[i\omega_{\textrm{ph}} b^{\dagger}_{(0)}+i\lambda (\sigma^{\rm L}_z+\sigma^{\rm R}_z)\big]-i\lambda (\sigma^{\rm L}_z+\sigma^{\rm R}_z)}{i\omega_{\textrm{ph}}}
\end{split}
\end{equation}

Using the same way we can get

$$b_{(t)}=\frac{e^{-i\omega_{\textrm{ph}} t}\big[-i\omega_{\textrm{ph}} b_{(0)}-i\lambda (\sigma^{\rm L}_z+\sigma^{\rm R}_z)\big]+i\lambda (\sigma^{\rm L}_z+\sigma^{\rm R}_z)}{-i\omega_{\textrm{ph}}}$$

Therefore

\begin{equation}
\begin{split}
b^{\dagger}_{(t)}b_{(t)}&=b^{\dagger}_{(0)}b_{(0)}+\frac{\lambda}{\omega_{\textrm{ph}}}(\sigma^{\rm L}_z+\sigma^{\rm R}_z)\\
&~~~\times\left[(1-e^{i\omega_{\textrm{ph}} t})b^{\dagger}_{(0)}+(1-e^{-i\omega_{\textrm{ph}} t})b_{(0)}\right]\\
&~~~+\frac{2\lambda^2 (\sigma^{\rm L}_z+\sigma^{\rm R}_z)^2}{\omega_{\textrm{ph}} ^2}[1-\cos(\omega_{\textrm{ph}} t)]
\end{split}
\end{equation}

Finally we can calculate the average number of phonon with respect to time.

\begin{equation}
\begin{split}
N_{\textrm{ph}} (t)&=\langle\Phi_{(0)}|b^{\dagger}_{(t)}b_{(t)}|\Phi_{(0)}\rangle \\
&=\langle\Phi_{(0)}|b^{\dagger}_{(0)}b_{(0)}+\frac{\lambda}{\omega_{\textrm{ph}}}(\sigma^{\rm L}_z+\sigma^{\rm R}_z)\big[(1-e^{i\omega_{\textrm{ph}} t})b^{\dagger}_{(0)}\\
&~~~+(1-e^{-i\omega_{\textrm{ph}} t})b_{(0)}\big]\\
&~~~+\frac{2\lambda^2 (\sigma^{\rm L}_z+\sigma^{\rm R}_z)^2}{\omega_{\textrm{ph}} ^2}[1-\cos(\omega_{\textrm{ph}} t)]|\Phi_{(0)}\rangle\\
&=\langle\downarrow^{\rm L},\downarrow^{\rm R},0|b^{\dagger}_{(0)}b_{(0)}+\frac{\lambda}{\omega_{\textrm{ph}}}(\sigma^{\rm L}_z+\sigma^{\rm R}_z)\big[(1-e^{i\omega_{\textrm{ph}} t})b^{\dagger}_{(0)}\\
&~~~+(1-e^{-i\omega_{\textrm{ph}} t})b_{(0)}\big]\\
&~~~+\frac{2\lambda^2 (\sigma^{\rm L}_z+\sigma^{\rm R}_z)^2}{\omega_{\textrm{ph}} ^2}[1-\cos(\omega_{\textrm{ph}} t)]|\downarrow^{\rm L},\downarrow^{\rm R}, 0\rangle\\
&=\frac{2\lambda^2}{\omega_{\textrm{ph}} ^2}[1-\cos(\omega_{\textrm{ph}} t)]\langle\downarrow^{\rm L}, \downarrow^{\rm R}, 0|(\sigma^{\rm L}_z+\sigma^{\rm R}_z)^2\\
&~~~|\downarrow^{\rm L}, \downarrow^{\rm R}, 0\rangle\\
&=\frac{8\lambda^2}{\omega_{\textrm{ph}} ^2}[1-\cos(\omega_{\textrm{ph}} t)]
\end{split}
\end{equation}

\medskip




%

\end{document}